\documentclass[aps,prl,preprint,groupedaddress]{revtex4}

\begin{document}


\title{Randall-Sundrum Model in the Presence of a Brane Bulk Viscosity }


\author{I. Brevik}
\email{iver.h.brevik@mtf.ntnu.no}
\author{A. Hallanger}

\affiliation{Department of Energy and Process Engineering, Norwegian University of Science and Technology, N-7491 Trondheim, Norway}


\date{\today}

\begin{abstract}
The presence of a bulk viscosity for the cosmic fluid on a single Randall-Sundrum brane is considered. The spatial curvature is assumed to be zero. The five-dimensional Friedmann equation is derived, together with the energy conservation equation for the viscous fluid. These governing equations are solved for some special cases: (i) in the low-energy limit when the matter energy density is small compared with brane tension; (ii) for a matter-dominated universe, and (iii) for a radiation-dominated universe. Rough numerical estimates, for the extreme case when the universe is at its Planck time, indicate that the viscous effect can be significant.
\end{abstract}


\maketitle

\section{I. Introduction}

As is known, the proposal of Randall and Sundrum (RS) \cite{randall99} to solve the hierarchy problem - in contrast to the approach of Arkani-Hamed {\it et al.} \cite{arkani98}, for instance - was to introduce a warped metric in five-dimensional space, the fifth dimension $y$ being compactified on an orbifold $S^1/Z_2$ of radius $R$ such that $-\pi R \le y \le \pi R$. In the two-brane model (RS1) the fixed points $y=0$ and $y=\pi R$ were the locations of the two three-branes, whereas in the one-brane model (RS2) the brane was taken to be located at $y=0$.

In a more realistic physical setting where the cosmic fluid is assumed to be present on the brane, the original RS warped metric has to be generalized. In addition to the brane tension (or four-dimensional cosmological constant), which we will call $\sigma$, one has to allow for  a fluid pressure $p$ and a mass-energy density $\rho$ in physical space. The form of metric that we shall consider in the following to describe this situation, is
\begin{equation}
ds^2= -n^2(t,y)dt^2+a^2(t,y)\delta_{ij}dx^idx^j +dy^2,
\label{1}
\end{equation}
where $n(t,y)$ and $a(t,y)$ are functions to be determined from Einstein's equations. This is a kind of metric that has been made use of earlier - cf. Refs.~[3-6], for instance - but we shall here specialize to the case of zero spatial curvature, $k=0$. There is by now ample experimental evidence supporting the flat space assumption.

We will in the following analyze a single flat membrane situated at the position $y=0$ in such a flat space. An isotropic viscous fluid with bulk viscosity $\zeta$ is taken to be present on the brane. Now, there are in general two different viscosity coefficients for a fluid, but we will henceforth ignore the second component, i.e., the shear viscosity $\eta$. The reason for this is the assumed perfect isotropy of the fluid. Actually this is a rather delicate point, due to the much larger magnitude of $\eta$ than of $\zeta$. Thus if one considers the conventional plasma epoch in the history of the universe a few tenths of seconds after the big bang, when the viscosity coefficients are calculable from ordinary kinetic theory, then one finds that $\eta$ is larger than $\zeta$ by more than 15 orders of magnitude. Even a small amount of anisotropy in the fluid would thus easily outweigh the effect of the bulk viscosity. This point is discussed more closely in Refs.~\cite{brevik94} and \cite{brevik00}.

The case of a bulk-viscous cosmic fluid has to our knowledge not been much studied studied before in a five-dimensional context, although we have recently become aware of the recent papers of Harko et al. in this direction \cite{chen01,harko03}. 
Some of our own motivations for undertaking this kind of study were the following:

First, there is the obvious need of generalization of the theory. The case of a nonviscous fluid is from a hydrodynamical point of view a very idealized situation, although it is a useful approximation in many cases.  A condition for five-dimensional viscous cosmology to have physical meaning, is that it is robust enough to encompass the presence of viscosity coefficients. Introduction of these coefficients implies that we are accounting for thermal irreversibility to the first order in the fluid velocity's derivatives.

Our second point relates to entropy. The nondimensional entropy per baryon in the universe is known to be very large, of the order of $10^9$ \cite{weinberg71}. Although this entropy is basically to be associated with microscopical processes one would expect, by analogy with ordinary hydrodynamics, that it is describable in terms of phenomenological parameters such as viscosity.

Finally, there are the energy conditions, the most important of which is probably the weak energy condition (WEC), $\rho +p \ge 0$ \cite{carter02}. The extent to which the presence of viscosity influences the energy  conditions seems to be to a large extent unexplored. Some discussion in the case of anisotropic cosmology can be found in \cite{brevik97}. The energy conditions are likely to be of importance also for the recent theories about phantom matter \cite{nojiri03}.

\section{II. Basic Concepts}

We summarize our assumptions: The metric, as mentioned, is given by Eq.~(\ref{1}), describing the space on the brane itself ($y=0)$ as well as in the surrounding bulk, the latter being empty except from the presence of a five-dimensional cosmological constant $\Lambda$. On the brane, the tension $\sigma$ is also assumed to be constant. The metric (\ref{1}) extends out to the horizon, determined by $g_{00}=0$. The five-dimensional Einstein equations are
\begin{equation}
R_{M N}-\frac{1}{2}g_{MN}R +g_{MN}\Lambda=\kappa^2T_{MN},
\label{2}
\end{equation}
where capital Latin indices denote five-dimensional indices $(t,x^1,x^2,x^3,y)$, and $\kappa^2=8\pi G_5$ is the five-dimensional gravitational coupling. The energy-momentum tensor describes every classical source except from the ``vacuum" fluid, i.e., $\Lambda$.

The components of the Ricci tensor are conveniently found using the Cartan formalism. For reference purposes we give them here:
\begin{equation}
R_{\hat{t}\hat{t}}=3\left( \frac{a'n'}{an}+\frac{\dot{a}\dot{n}}{an^3}-\frac{\ddot{a}}{an^2} \right)+\frac{n''}{n},
\label{3}
\end{equation}
\begin{equation}
R_{\hat{i}\hat{i}}=\frac{\ddot{a}}{an^2}-\frac{\dot{a}\dot{n}}{an^3}-\frac{a'n'}{an}+2\left[ \left(\frac{\dot{a}}{an}\right)^2
-\left(\frac{a'}{a}\right)^2\right]-\frac{a''}{a}\quad \rm{(no\; sum)},
\label{4}
\end{equation}
\begin{equation}
R_{\hat{y}\hat{y}}=-\frac{n''}{n}-\frac{3a''}{a},
\label{5}
\end{equation}
\begin{equation}
R_{\hat{t}\hat{y}}=3\left( \frac{\dot{a}n'}{an^2}-\frac{\dot{a}'}{an}\right).
\label{6}
\end{equation}
Since the relationship between the components $T_{\hat{M}\hat{N}}$ of the energy-momentum tensor in orthonormal basis and the components $T_{MN}$ in coordinate basis are
\begin{equation}
T_{tt}=n^2 T_{\hat{t}\hat{t}},
\quad T_{ij}=a^2T_{\hat{i}\hat{j}},\quad T_{ty}=nT_{\hat{t}\hat{y}},\quad T_{yy}=T_{\hat{y}\hat{y}},
\label{7}
\end{equation}
we get from Eq.~(\ref{2})
\begin{equation}
3\left\{ \left( \frac{\dot{a}}{a}\right)^2 
-n^2 \left[\frac{a''}{a}
+\left(\frac{a'}{a} \right)^2 \right]\right\}-\Lambda n^2=\kappa^2 T_{tt},
\label{8}
\end{equation}
\begin{equation}
\delta_{ij} \Bigg\{ a^2\Big[\frac{a'}{a}\left(\frac{ a'}{a}+\frac{2n'}{n}\right) +\frac{2a''}{a}+\frac{n''}{n} \Big] 
+\frac{a^2}{n^2}\Big[ \frac{\dot{a}}{a}\left(-\frac{\dot{a}}{a}+\frac{2\dot{n}}{n}\right)-\frac{2\ddot{a}}{a}\Big] +
\Lambda a^2 \Bigg\}
=\kappa^2 T_{ij},
\label{9}
\end{equation}
\begin{equation}
3\left( \frac{\dot{a}}{a}\frac{n'}{n}-\frac{\dot{a}'}{a} \right)=\kappa^2T_{ty},
\label{10}
\end{equation}
\begin{equation}
\frac{3a'}{a}\left(\frac{a'}{a}+\frac{n'}{n}\right)-\frac{3}{n^2}\left[\frac{\dot{a}}{a}\left( \frac{\dot{a}}{a}-\frac{\dot{n}}{n}\right)
+\frac{\ddot{a}}{a}\right]+\Lambda=\kappa^2 T_{yy}.
\label{11}
\end{equation}
Here $a=a(t,y),\, n=n(t,y)$, overdots and primes meaning derivatives with respect to $t$ and $y$ repectively. So far, the form of the energy-momentum tensor has not been specified.

A note on dimensions: As the spatial curvature $k=0$, it is convenient to let the dimensions be carried by the coordinates so that $a$ and $n$ become nondimensional. Thus $[a]=[n]=1,\,[x^\mu]=[y]={\rm cm},\,[\Lambda]={\rm cm}^{-2},\, [\kappa^2]={\rm cm^3}$.

\section{III. Energy-Momentum Tensor and Governing Equations}

The governing equations for the metric components are obtained by integrating Einstein's equations across the singular boundary at $y=0$. As for the energy-momentum tensor, the contribution to it from the constant brane tension is $T_{MN}=-\delta(y) \sigma\, \delta_M^\mu\delta_N^\nu\, g_{\mu\nu}$. In an orthonormal frame this means that $T_{\hat{t}\hat{t}}=\delta(y)\sigma$, $T_{\hat{i}\hat{j}}=-\delta(y)\sigma$, which effectively implies $p=-\sigma$ on the brane, i.e., the conventional pressure-energy relationship for a domain wall \cite{vilenkin81}.

Adding the energy-momentum tensor $T_{\mu\nu}^{\rm fluid}$ for the viscous fluid, we get
\begin{equation}
T_{MN}=\delta(y)(-\sigma g_{\mu\nu}+T_{\mu\nu}^{ \rm fluid})\delta_M^\mu\delta_N^\nu,
\label{12}
\end{equation}
where we insert the conventional four-dimensional expression
\begin{equation}
T_{\mu\nu}^{\rm fluid}=\rho U_\mu U_\nu+(p-\zeta \theta)h_{\mu\nu}.
\label{13}
\end{equation}
Here $U^\mu=(U^0,U^i)$ is the fluid's four-velocity, $\theta={U^\mu}_{;\mu}$ is the scalar expansion, and $h_{\mu\nu}=g_{\mu\nu}+U_\mu U_\nu$ is the projection tensor. We work henceforth in an orthonormal frame, corresponding to $U^\mu=(1,0,0,0).$ With the notation $a_0(t)=a(t,y=0)$ and $n_0(t)=n(t,y=0)$ we can write the metric on the brane as
\begin{equation}
ds^2=-n_0^2(t)dt^2+a_0^2(t)\delta_{ij}dx^idx^j.
\label{14}
\end{equation}
The scalar expansion is calculated as $\theta=3\dot{a}_0/a_0+\dot{n}_0/n_0$. We impose the gauge condition $n_0(t)=1$ which physically means that the proper time on the brane is taken as the time coordinate, and so get
\begin{equation}
\theta=\frac{3\dot{a}_0}{a_0} \equiv 3H_0.
\label{15}
\end{equation}
This is the same relationship as in ordinary FRW cosmology. Letting $\tilde{p}$ denote the effective pressure,
\begin{equation}
\tilde{p}=p-3H_0\zeta,
\label{16}
\end{equation}
we thus have
\begin{equation}
T_{tt}^{\rm fluid}=\rho, \quad T_{ij}^{\rm fluid}=\tilde{p}a_0^2\delta_{ij}.
\label{17}
\end{equation}
Consider next the junction conditions. The metric is continuous across the brane, but its derivatives are not. The distributional parts of $a''$ and $n''$ have to be matched with the distributional parts of the energy-momentum tensor. Following Ref.~\cite{binetruy00} we write $a''=\hat{a}''+[a']\delta(y)$, where $[a']=a'(y=0^+)-a'(y=0^-)$ is the jump across $y=0$ and $\hat{a}''$ is the nondistributional part. Similarly, we write $n''= \hat{n}''+[n']\delta(y)$ . The matching procedure applied to Eqs.~(\ref{8}) and (\ref{9}) yields
\begin{equation}
\frac{[a']}{a_0}=-\frac{1}{3}\kappa^2(\sigma+\rho),
\label{18}
\end{equation}
\begin{equation}
[n']=-\frac{1}{3}\kappa^2\sigma+\frac{2}{3}\kappa^2\rho+\kappa^2\tilde{p}
\label{19}
\end{equation}
for the distributional parts, and
\begin{equation}
\left( \frac{\dot{a}}{na}\right)^2-\frac{a''}{a}-\left(\frac{a'}{a}\right)^2=\frac{1}{3}\Lambda,
\label{I}
\end{equation}
\[ \frac{a'}{a}\left( \frac{a'}{a}+\frac{2n'}{n}\right)+\frac{2a''}{a}+\frac{n''}{n} \]
\begin{equation}
+\frac{1}{n^2}\Big[ \frac{\dot{a}}{a} \left( -\frac{\dot{a}}{a}+\frac{2\dot{n}}{n} \right)
-\frac{2\ddot{a}}{a}\Big] = -\Lambda
\label{II}
\end{equation}
for the nondistributional parts. (For simplicity we have written $a''$ and $n''$ instead of $\hat{a}''$ and $\hat{n}''$ in Eqs.~(\ref{I}) and (\ref{II}).) In the junction conditions (\ref{18}) and (\ref{19}) we have made use of $n_0=1.$ Equation (\ref{18}) is seen to be formally the same as for an ideal fluid, whereas Eq.~(\ref{19}) is explicitly viscosity dependent.

We assume henceforth that there is no flux of energy in the $y$ direction. It means,
\begin{equation}
T_{ty}=0.
\label{20}
\end{equation}
Equation (\ref{10}) thus yields
\begin{equation}
\frac{\dot{a}'}{a}=\frac{\dot{a}}{a}\,\frac{n'}{n},
\label{III}
\end{equation}
or by integration
\begin{equation}
n(t,y)=\frac{\dot{a}(t,y)}{\dot{a}_0(t)},
\label{21}
\end{equation}
for arbitrary $y$. 

By means of Eq.~(\ref{III}) we can write Eq.~(\ref{I}) as
\begin{equation}
\frac{d}{dy}\Big[ (aa')^2-\left(\frac{\dot{a}a}{n}\right)^2\Big]=-\frac{2}{3}\Lambda
\frac{d}{dy}[\frac{1}{4}a^4].
\label{IV}
\end{equation}
Integrating this equation with respect to $y$ we get
\begin{equation}
\left( \frac{\dot{a}}{na}\right)^2=
\frac{1}{6}\Lambda+\left( \frac{a'}{a}\right)^2+\frac{C}{a^4},
\label{V}
\end{equation}
where $C=C(t)$ is an integration constant.

Assuming $Z_2$ symmetry $y \rightarrow -y$ for the scale factor $a$ across the brane as well as continuity for $y\rightarrow 0$, we obtain from the boundary condition (\ref{18})
\begin{equation}
\frac{a_0'}{a_0}=-\frac{1}{6}\kappa^2\,(\sigma+\rho),
\label{VI}
\end{equation}
at $y=0$. Also by evaluating Eq.~(26) on the brane and removing the $y$ derivatives by inserting Eq.~(\ref{VI}) we finally obtain (remembering $n_0(t)=1$)  
\begin{equation}
H_0^2=\lambda+\frac{1}{18}\kappa^4\sigma \rho+\frac{1}{36}\kappa^4\rho^2+\frac{C}{a_0^4}
\label{22}
\end{equation}
 (recall that subscript zero refers to the brane). We have here introduced the quantity
\begin{equation}
\lambda=\frac{1}{6}\Lambda+\frac{1}{36}\kappa^4\sigma^2,
\label{23}
\end{equation}
which can be interpreted as the effective four-dimensional cosmological constant in the five-dimensional theory.  The ``new" Friedmann equation (\ref{22}) can be contrasted with the conventional Friedmann equation in four-dimensional cosmology (when $k=0$):
\begin{equation}H_0^2=\frac{1}{3}\Lambda_4+\frac{1}{3}\kappa_4^2\,\rho,
\label{24}
\end{equation}
where $\Lambda_4$ and $\kappa_4^2=8\pi G_4$ are four-dimensional quantities. The essential new properties of Eq.~(\ref{22}) are thus the presence of a $\rho^2$-term, and also the presence of  a ``radiation" term  $C/a_0^4$. 

The (nondistributional) Einstein equation (\ref{I}) can be solved analytically for $a(t,y)$. Inserting Eq.~(\ref{21}) into Eq.~(\ref{I}) and multiplying with $a^2$ we obtain
\begin{equation}
(aa')'+\frac{1}{3}\Lambda \,a^2-\dot{a}_0^2=0.
\label{VII}
\end{equation} 
Let us first assume that $\Lambda>0$, i.e., the de Sitter (dS) case, and let us take as trial solution
\begin{equation}
a^2(t,y)=A\cos(2\mu_d\,y)+B\sin(2\mu_d\,y)+D,
\label{VIII}
\end{equation}
where $A,B,D$ are arbitrary functions of time, and where $\mu_d \equiv \sqrt{\Lambda/6}$. In order for Eq.~(\ref{VIII}) to satisfy Eq.~(\ref{VII}), we must take
\begin{equation}
D=\frac{3}{\Lambda}\dot{a}_0^2=\frac{1}{2}a_0^2\Big[1+\frac{\kappa^4}{6\Lambda}(\sigma+\rho)^2\Big]+\frac{3C}{\Lambda a_0^2},
\label{IX}
\end{equation}
where the elimination of $\dot{a}_0$ in the final step is done by means of Eq.~(\ref{22}).

The functions $A$ and $B$ in Eq.~(\ref{VIII}) are determined by using the assumption of $Z_2$ symmetry, together with the junction condition (\ref{18}). Writing $A^+ \equiv A|_{y=0^+},\; A^- \equiv A|_{y=0^-},\; B^+ \equiv B|_{y=0^+}$ and $B^- \equiv B|_{y=0^-}$, and using simple symmetry relations for the trigonometric functions, we obtain from Eq.~(\ref{VIII})
\begin{equation}
A^+ =A^- \equiv \bar{A},\quad B^+=-B^-\equiv \bar{B}.
\label{X}
\end{equation}
After a simple calculation using Eqs.~(\ref{18}), (\ref{VIII}), (\ref{IX}), and (\ref{X}), we find 
\begin{equation}
\bar{A}=\frac{1}{2}a_0^2\Big[ 1-\frac{\kappa^4}{6\Lambda}(\sigma +\rho)^2 \Big]-\frac{3C}{\Lambda\,a_0^2},
\label{XI}
\end{equation}
and
\begin{equation}
\bar{B}=-\frac{\kappa^2}{6\mu_d}(\sigma+\rho)a_0^2.
\label{XII}
\end{equation}
The full expression for $a(t,y)$ is thus
\[ a^2(t,y)=\frac{1}{2}a_0^2\Big[ 1+\frac{\kappa^4}{6\Lambda}(\sigma+\rho)^2\Big]+\frac{3C}{\Lambda \,a_0^2} \]
\[ +\left\{ \frac{1}{2}a_0^2\Big[1-\frac{\kappa^4}{6\Lambda}(\sigma+\rho)^2\Big]-\frac{3C}{\Lambda\,a_0^2}\right\}
\cos(2\mu_d\,y) \]
\begin{equation}
-\frac{\kappa^2}{6\mu_d}(\sigma+\rho)a_0^2\sin(2\mu_d\,|y|).
\label{XIII}
\end{equation}
The corresponding solution of Einstein's equation for an AdS bulk space (i.e., $\Lambda<0$) is
\[ a^2(t,y)=\frac{1}{2}a_0^2\Big[ 1+\frac{\kappa^4}{6\Lambda}(\sigma+\rho)^2\Big]+\frac{3C}{\Lambda \,a_0^2} \]
\[ +\left\{ \frac{1}{2}a_0^2\Big[1-\frac{\kappa^4}{6\Lambda}(\sigma+\rho)^2\Big]-\frac{3C}{\Lambda\,a_0^2}\right\}
\cosh(2\mu\,y) \]
\begin{equation}
-\frac{\kappa^2}{6\mu}(\sigma+\rho)a_0^2\sinh(2\mu\,|y|),
\label{XIV}
\end{equation}
where $\mu \equiv \sqrt{-\Lambda/6}$.
Equations (\ref{XIII}) and (\ref{XIV}) are the same solutions as given in Ref.~\cite{brevik02}, but here with a fluid on the brane (i.e., $\sigma \rightarrow \sigma+ \rho$).

The second junction condition, Eq.~(\ref{19}), relates to the conservation equation for energy. Using the same trial solution as before, i.e., Eq.~(\ref{VIII}) for a dS bulk space (which we know satifies the Einstein equation as long as $D$ is given by Eq.~(\ref{IX})), and taking into account the relation
\begin{equation}
n'(t,y)=\frac{\dot{a}'(t,y)}{\dot{a}_0(t)},
\label{25}
\end{equation}
which follows from Eq.~(\ref{21}) by differentation, we find by evaluating $n'$ on both sides of the brane
\begin{equation}
\dot{\rho}+3(\rho+p)H_0-9\zeta H_0^2=0.
\label{26}
\end{equation}
The calculation is analogous to that leading to Eq.~(\ref{XIII}), but now with the junction condition for $n$, Eq.~(\ref{19}), taken into account.

Equation (\ref{26}) is the same as that obtained in ordinary four-dimensional cosmology \cite{gron90,brevik94,brevik02a}. This is somewhat surprising, as there is seemingly no simple physical reason why it should be so.

The anti-de Sitter (AdS) universe, $\Lambda <0$, yields the same result.

\section{IV. Cosmological Solutions}

We introduce the equation of state in the conventional form
\begin{equation}
p=(\gamma-1)\rho,
\label{27}
\end{equation}
where $\gamma$ is a constant lying in the interval $0\le \gamma \le 2$. As from now on we will consider cosmological solutions only, we will henceforth leave out subscripts zero referring to the brane position $y=0$.

As is commonly believed, the evolution of our universe consists of four epochs: (1) the early inflationary epoch, (2) the radiation dominated epoch, (3) the matter dominated epoch, and (4) the present mini-inflationary epoch (cf. for instance, the discussion in \cite{brevik02}). The energy density of our universe is dominated by the effective cosmological constant in epochs (1) and (4). We will assume that the influence of the radiation term $C/a^4$ is negligible except in the radiation era, epoch (2).

With $C=0$ the governing equations reduce to the set
\begin{equation}
H^2=\lambda+\frac{1}{18}\kappa^4\sigma \rho+\frac{1}{36}\kappa^4\rho^2,
\label{28}
\end{equation}
\begin{equation}
\dot{\rho}+\gamma \rho \,\theta-\zeta \theta^2=0,
\label{29}
\end{equation}
with $\theta=3H$.

Before embarking on the viscous case, let us briefly consider the situation when $\zeta=0$.

\subsection{A. Nonviscous fluid}

From Eqs.~(\ref{28}) and (\ref{29}) we get for the cosmological time
\begin{equation}
t=-\frac{2}{\gamma \kappa^2\sigma}\int \frac{dx}{x\sqrt{x^2+2x+36\lambda /\kappa^4\sigma^2}},
\label{30}
\end{equation}
where $x=\rho/\sigma$ is a nondimensional parameter. From Eq.~(\ref{29}) it follows that when $\zeta=0$ the following constant emerges:
\begin{equation}
\rho a^{3\gamma}=\rho_* a_* ^{3\gamma},
\label{31}
\end{equation}
the asterisk meaning a reference time, so far unspecified.

If $\lambda=0$ we have the solution
\begin{equation}
a^{3\gamma}=\frac{1}{2}\rho_*a_*^{3\gamma}\,\gamma \kappa^2t\left(1+\frac{1}{4}\gamma \kappa^2\sigma t\right),
\label{32}
\end{equation}
whereas if $\lambda>0$, 
\begin{equation}
a^{3\gamma}=\rho_* a_*^{3\gamma}\left[ \frac{\kappa^4\sigma}{18\lambda}\sinh^2\frac{t}{t_\lambda}+\frac{\kappa^2}{6\sqrt{\lambda}}
\sinh \frac{2t}{t_\lambda}\right],
\label{33}
\end{equation}
with $t_\lambda=2/(3\gamma \sqrt{\lambda}).$ The initial condition is $a(t=0)=0$.

Equations (\ref{32}) and (\ref{33}) are in agreement with Binetruy {\it et al.} \cite{binetruy00a}. The solution for the case $\lambda<0$ was however not given in \cite{binetruy00a}, so we give it here:
\begin{equation}
a^{3\gamma}=\rho_*a_*^{3\gamma}\left[-\frac{\kappa^4\sigma}{18\lambda}\sin^2\frac{t}{t_\lambda}+\frac{\kappa^2}{6\sqrt{-\lambda}}\sin 
\frac {2t}{t_\lambda}\right],
\label{34}
\end{equation}
where now $t_\lambda=2/(3\gamma\sqrt{|\lambda|})$.

It is of interest to compare with four-dimensional cosmology. The discovery that the cosmic expansion is accelerating has replaced the Einstein - de Sitter model with the Friedmann - Lema\^{i}tre model as the standard four-dimensional model \cite{gron02}. This is a flat universe with cosmological constant $\Lambda_4$ and pressure-free matter ($\gamma=1$). According to this model,
\begin{equation}
a^3= \rho_*a^3_*\,\frac{\kappa_4^2}{\Lambda_4}\,\sinh^2 \frac{t}{t_\Lambda},
\label{35}
\end{equation}
with $t_\Lambda=2/\sqrt{3\Lambda_4}$. Equation (\ref{35}) is seen to correspond to the {\it first} term in Eq.~(\ref{33}) in view of the relationships
\begin{equation}
\lambda \rightarrow \frac{1}{3}\Lambda_4,\quad \kappa^4\sigma\rightarrow 6\kappa_4^2,\quad t_\lambda\rightarrow t_\Lambda,
\label{36}
\end{equation}
the two first of which follow from a comparison of Eqs.~(\ref{22}) and (\ref{24}).

\subsection{B. Low energy limit}

If $\rho\ll \sigma$, i.e., if the matter density in the universe is much smaller than the intrinsic cosmological constant, the Friedmann equation (\ref{22}) reduces to
\begin{equation}
H^2=\lambda+\frac{1}{18}\kappa^4\sigma \rho,
\label{37}
\end{equation}
which is is seen to reduce to the four-dimensional Friedmann equation, Eq.~(\ref{24}), in view of the relationships given in Eq.~(\ref{36}). In this case the solutions of five-dimensional cosmology can thus be found from the conventional four-dimensional theory.

Taking the time derivative of Eq.~(\ref{37}) and using the energy conservation equation (\ref{29}), we get the following equation for the scalar expansion ($\theta=3H$):
\begin{equation}
\dot{\theta}(t)+\frac{1}{2}\gamma \theta^2(t)-\frac{1}{4}\kappa^4\,\sigma \,\zeta \,\theta(t)-\frac{9}{2}\gamma \lambda=0.
\label{38}
\end{equation}
If $\lambda=0$, in view of $\kappa^4\sigma \rightarrow 6\kappa_4^2$ this equation is the same as Eq.~(31) in Ref.~\cite{brevik94}. Now assume that  $\lambda>0$. Integration of Eq.~(\ref{38}) yields
\begin{equation}
t=\frac{2}{\gamma \sqrt{K}}\ln \Big| \frac{2\theta-\beta+\sqrt{K}}{2\theta-\beta-\sqrt{K}} \Big|,
\label{39}
\end{equation}
where $\beta=\kappa^4\sigma\zeta/(2\gamma)$ and $K=36\lambda+\beta^2$. Inverting this expression we get
\begin{equation}
\theta=\frac{1}{2}\beta+\frac{1}{2}\sqrt{K}\coth \left( \frac{1}{4}\gamma\sqrt{K}\,t\right).
\label{40}
\end{equation}
Near $t=0$ the viscosity is not expected to be important. Ignoring $\beta$ we obtain $\theta \rightarrow 2/(\gamma t)$ when $t\rightarrow 0$. Thus when $\rho \ll \sigma$, five-dimensional cosmology predicts the scalar expansion to depend on the form of the equation of state near $t=0$. This behaviour of $\theta$ is the same as found in four-dimensional cosmology. In the latter case, for any value of the curvature parameter $k$, one has  $a \propto t^{2/(3\gamma)}$ for $t\rightarrow 0$, resulting in $H=2/(3\gamma t)$, equivalent to the expression for $\theta$ given above.

\subsection{ C. Matter-dominated universe, $\sigma=0$}
\label{IV.C}

In conventional four-dimensional cosmology one assumes, as already mentioned, that one epoch was dominated by a non-accelerating expansion in which cold matter dominated the universe. This is the Einstein - de Sitter model: a flat universe with no cosmological constant.

In five-dimensional cosmology the Friedmann equation (\ref{28}) in a matter-dominated universe would describe a universe in which the $\rho^2$ term dominates over the $\sigma \rho$ term. We will in the present section simply assume that the brane tension $\sigma$ is zero. Note that the effective cosmological constant $\lambda$ of Eq.~(\ref{23}) can still dominate over the $\rho^2$ term if $\Lambda$ is large enough, thus possibly describing an inflationary universe.

The Einstein - de Sitter model corresponded to a vanishing pressure $p$, i.e., $\gamma=1$. Here, we will admit that there exists a matter pressure, so that $\gamma$ will be kept unspecified. As we are still outside the radiation epoch we assume, as before, that $C=0$.

The Friedmann equation becomes now
\begin{equation}
H^2=\lambda+\frac{1}{36}\kappa^4\rho^2
\label{41}
\end{equation}
which, together with the energy conservation equation (\ref{29}), yields the following equation for the scalar expansion:
\begin{equation}
\dot{\theta}+\gamma(\theta^2-9\lambda)-\frac{1}{2}\kappa^2\zeta \theta \sqrt{\theta^2-9\lambda}=0.
\label{42}
\end{equation}
We first consider the de Sitter case, $\Lambda>0$, which implies that also $\lambda>0$ in view of the assumption $\sigma=0$. With the substitution $\theta=3\sqrt{\lambda} \cosh x$ we obtain
\begin{equation}
-3\gamma dt=\frac{dx}{\sinh x-\delta \cosh x},
\label{43}
\end{equation}
with $\delta=\kappa^2\zeta/(2\gamma)$. Integrating this equation we obtain a relation between $t$ and $x$. The resulting expression for $\theta$ will depend on the magnitude of the viscosity. Assume first that $\kappa^2\zeta<2\gamma$. We then obtain
\begin{equation}
\theta=\frac{3}{2}\sqrt{\lambda}\left\{ \frac{1}{\hat{B}}\tanh \left(\frac{3}{2}\gamma \sqrt{\lambda}\,\hat{\beta}\,t\right)
+\hat{B}\coth \left(\frac{3}{2}\gamma \sqrt{\lambda}\,\hat{\beta}\,t\right) \right\},
\label{44}
\end{equation}
where
\begin{equation}
\hat{B}=\sqrt{\frac{2\gamma +\kappa^2\zeta}{2\gamma-\kappa^2\zeta}},\quad \hat{\beta}=\sqrt{1-\left(\frac{\kappa^2\zeta}{2\gamma}\right)^2}.
\label{45}
\end{equation}
The form of Eq.~(\ref{44}) implies that $\theta\rightarrow \hat{B}/(\gamma \hat{\beta}t)$ when $t\rightarrow 0$.

If $\kappa^2\zeta>2\gamma$ we get
\begin{equation}
\theta=\frac{3}{2}\sqrt{\lambda}\left\{ \frac{1}{\tilde{B}}\tan \left(\frac{3}{2}\gamma \sqrt{\lambda}\,\tilde{\beta}\,t\right)
+\tilde{B}\cot \left(\frac{3}{2}\gamma \sqrt{\lambda}\,\tilde{\beta}\,t\right)\right\},
\label{46}
\end{equation}
where
\begin{equation}
\tilde{B}=\sqrt{\frac{\kappa^2\zeta+2\gamma}{\kappa^2\zeta-2\gamma}},\quad \tilde{\beta}=\sqrt{\left(\frac{\kappa^2\zeta}{2\gamma}
\right)^2-1}.
\label{47}
\end{equation}
For an AdS universe, $\Lambda<0$ implying $\lambda<0$, we obtain by an analogous calculation
\begin{equation}
\theta=\frac{3}{2}\sqrt{-\lambda}\left\{ -\frac{1}{\hat{B}}\tan\left(\frac{3}{2}\gamma\sqrt{-\lambda}\,\hat{\beta}\,t\right)
+\hat{B}\cot\left(\frac{3}{2}\gamma\sqrt{-\lambda}\,\hat{\beta}\,t\right)\right\}
\label{48}
\end{equation}
when $\kappa^2\zeta<2\gamma$, and
\begin{equation}
\theta=\frac{3}{2}\sqrt{-\lambda}\left\{-\frac{1}{\tilde{B}}\tanh \left(\frac{3}{2}\gamma\sqrt{-\lambda}\,\tilde{\beta}\,t\right)
+\tilde{B}\coth\left(\frac{3}{2}\gamma\sqrt{-\lambda}\,\tilde{\beta}\,t\right)\right\},
\label{49}
\end{equation}
when $\kappa^2\zeta>2\gamma$. To our knowledge, these expressions have not been given before.

Once $\theta(t)$ is known, the energy density $\rho(t)$ is easily obtained from Eq.~(\ref{41}). The scale factor $a(t)$ itself is obtained by integrating the equations above. Let us give $a(t)$ for the case of a dS space with $\kappa^2\zeta<2\gamma$:
\begin{equation}
a^{3\gamma \hat{\beta}}=\left( 2\rho_*a_*^{3\gamma}\frac{\kappa^2}{3\sqrt{\lambda}}\right)^{\hat{\beta}} \cosh^{1/\hat{B}}\left(\frac{3}{2}\gamma
\sqrt{\lambda}\,\hat{\beta}\,t\right)\sinh^{\hat{B}}\left(\frac{3}{2}\gamma\sqrt{\lambda}\,\hat{\beta}\,t\right).
\label{50}
\end{equation}
In the nonviscous case corresponding to $\hat{B}=\hat{\beta}=1$, this expression reduces to
\begin{equation}
a^{3\gamma}=\rho_*a_*^{3\gamma}\frac{\kappa^2}{6\sqrt{\lambda}}\sinh \frac{2t}{t_\lambda}, \quad t_\lambda=\frac{2}{3\gamma\sqrt{\lambda}},
\label{51}
\end{equation}
in agreement with Eq.~(\ref{33}) (in which the first term is zero).

It is natural to ask: Which of the above options are most likely to describe real physics in the early universe? Most likely one would expect that the bulk viscosity is low, so that the inequality $\kappa^2\zeta<2\gamma$ is obeyed. This should support the validity of Eqs.~(\ref{44}) or (\ref{48}). However the situation may be more complicated, at least near the Planck time; as we will see more closely in subsection D it may happen that the contribution from the bulk viscosity becomes quite significant. Moreover, it follows of course mathematically that if $\gamma$ happens to be close to zero in some epoch in the universe's history, thus corresponding to an approximate ``vacuum fluid", $p=-\rho$, then $\kappa^2\zeta$ can easily exceed $2\gamma$.

\subsection{D. Radiation-dominated universe}
\label{IV.D}

In the epoch of the universe's history where radiation dominates, it is no longer permissible to neglect the term $C/a^4$ in Friedmann's equation. The equation of state corresponding to $\gamma=4/3$ is
\begin{equation}
p=\frac{1}{3}\rho,
\label{52}
\end{equation}
and the governing equations are
\begin{equation}
H^2=\lambda+\frac{1}{18}\kappa^4\sigma\rho+\frac{1}{36}\kappa^4\rho^2+\frac{C}{a^4},
\label{53}
\end{equation}
\begin{equation}
\dot{\rho}+\frac{4}{3}\rho\,\theta-\zeta\theta^2=0.
\label{54}
\end{equation}
As our primary aim is to study the influence from viscosity, we shall in this subsection ignore brane tension and set $\sigma=0$.

To parametrize the viscosity, we make the following ansatz for the ``kinematic" viscosity coefficient $\zeta/\rho$ associated with   $\zeta$:
\begin{equation}
\frac{\zeta}{\rho}=\frac{\alpha}{3\theta},
\label{55}
\end{equation}
where $\alpha$ is a constant (the factor 3 is for convenience). This choice is motivated partly by analytic tractability of the formalism. In  addition, as a more physical motivation, we note that it has some similarity with the ansatz made by Murphy three decades ago in his classic paper on viscous cosmology \cite{murphy72}. Murphy assumed that $\zeta/\rho=$ constant. Recalling from conventional cosmology that $\theta=3/(2t)$ when $t\rightarrow 0$, we see that our ansatz describes a weaker influence from viscosity near the big bang than does the ansatz of Murphy: Eq.~(\ref{55}) predicts the kinematic viscosity to start from zero at $t=0$ and thereafter increase linearly with $t$.

Inserting Eq.~(\ref{55}) into Eq.~(\ref{54}) we obtain
\begin{equation}
\rho a^{4-\alpha}=\rho_* a_*^{4-\alpha},
\label{56}
\end{equation}
where the asterisk as before means an unspecified instant. The presence of the coefficient $\alpha>0$ thus implies that $\rho$ decreases more slowly with increasing $a$ than what is the case for a nonviscous fluid.

The relationship between $t$ and $a$ can be written as an integral ($x=a^4)$:
\begin{equation}
\sqrt{\lambda}\,t=\frac{1}{4}\int_0^{a^4} \frac{dx}{\sqrt{x^2+(C/\lambda)x+(Q_\alpha/\lambda) x^{\alpha/2}}},
\label{57}
\end{equation}
where
\begin{equation}
 Q_\alpha=\left( \frac{\kappa^2}{6}\rho_*a_*^{4-\alpha}\right)^2.
\label{58}
\end{equation}
We have expressed the integrand in Eq.~(\ref{57}) in terms of nondimensional quantities (recall that $[a]=[x]=1$, so that $ [C]=[\lambda]=[Q_\alpha]={\rm cm^{-2}}$).

The integral in Eq.~(\ref{57}) is easily calculable in three special cases: $\alpha =0,2,4$. First, let us set $\alpha=0$, taking for definiteness $\lambda$ to be positive:
\begin{equation}
\sqrt{\lambda}\,t=\frac{1}{4}\ln 
\frac{ a^4 +(C/2\lambda)+\sqrt{a^8+(C/\lambda)a^4+Q_0/\lambda}}
{(C/2\lambda)+\sqrt{Q_0/\lambda}}.
\label{59}
\end{equation}
As usual, we have assumed $a=0$ for $t=0$. For $\alpha=2$, we can simply use the same expression (\ref{59}), with the substitutions
\begin{equation}
C \rightarrow C+ Q_2,\quad Q_0\rightarrow 0.
\label{60}
\end{equation}
The effect from viscosity is thus contained in $Q_2$. If $\alpha=4$, an analogous substitution can be done. This case is however pathological, since it corresponds to $\rho$= constant according to Eq.~(\ref{56}).

If $\lambda=0$, we have
\begin{equation}
t=\frac{1}{4}\int_0^{a^4} \frac{dx}{\sqrt{Cx+Q_\alpha x^{\alpha/2}}},
\label{61}
\end{equation}
giving for $\alpha=0$
\begin{equation}
t=\frac{1}{2C}\left(\sqrt{Ca^4+Q_0}-\sqrt{Q_0}\right).
\label{62}
\end{equation}
The viscous result for $\alpha=2$ is obtained by means of the same substitution as above, Eq.~(\ref{60}).

For general values of $\alpha$ the integral in Eq.~(\ref{57}) is of course easily done numerically. Most likely, the viscous effects that we consider here are to be attributed to the very early times in the history of the universe. As a rather extreme physical example, let us consider the evolution of the universe from $t=0$ up to the Planck time, $T_P=L_P=\sqrt{G_4}=5.4 \times 10^{-44}$ s (or $1.6 \times 10^{-33}$ cm), and let us identify the asterisk quantities in Eq.~(\ref{56}) with this particular instant. Normalizing the nondimensional scale factor such that $a_*=1$,  meaning that the value of $a(t)$ is given in Planck lengths, we may estimate the magnitude of the left hand side of Eq.~(\ref{57}) (with $t=t_*=T_P$) as follows: The magnitude of $\lambda$, equal to $\Lambda/6$ in our case since $\sigma=0$, is taken to be of the same order as the RS fine-tuning energy scale $k=\sqrt{|\Lambda|/6}$ (ignoring here the minus sign for $\Lambda$). Thus $\sqrt{\lambda} \sim k$. Moreover, we adopt the estimate $M_5 \sim M_4 \sim k$ which is customary in the RS context \cite{camera03}. As $M_4 =G_4^{-1/2}$ we then get $\sqrt{\lambda}\,T_P \sim 1$. Moreover, $Q_\alpha=Q_0$ in view of $a_*=1$. As very little seems to be known about the magnitudes of $C/\lambda$ and $Q_0/\lambda$ we shall here simply assume, as a working hypothesis, that  these quantities  are equal to unity. Thus we obtain from Eq.~(\ref{57}), when omitting the prefactor 1/4, 
\begin{equation}
1\sim \int_0^1 \frac{dx}{\sqrt{x^2+x+x^{\alpha/2}}}
\label{63}
\end{equation}
as a condition for determining the magnitude of $\alpha$. Due to the crudeness of the argument we cannot determine $\alpha$ quantitatively from this condition, but we find it of interest to note that Eq.~(\ref{63}) can be satisfied with $\alpha$ being of the order of unity.

Finally, let us estimate the magnitude of the viscosity coefficient. We have $\kappa^2 = 8\pi G_5= 8\pi /M_5^3 \sim 8\pi/M_4^3 \sim 10^{-97}\; {\rm cm}^3$ which, together with the estimate $Q_0/\lambda=\left( \frac{\kappa^2}{6}\rho_*\right)^2/\lambda \sim 1$, leads to $\rho_* \sim 10^{130}\;{\rm cm}^{-4}$ (or $10^{92}\; \rm g/cm^3$). With $\theta=3/(2t)$ we then obtain at Planck time, assuming $\alpha \sim 1$, 
\begin{equation}
\frac{\zeta_*}{ \rho_*}\sim 10^{-33}\; {\rm cm}, \quad \zeta_*\sim 10^{97} \; {\rm cm^{-3}}.
\label{64}
\end{equation}
Thus the bulk kinematic viscosity $\zeta_*/\rho_*$ is small, and the bulk viscosity $\zeta_*$ itself is large. Moreover, we get $\kappa^2\zeta_* \sim 1$. Somewhat remarkably, this indicates that we are in the transition region between the two cases $\kappa^2 \zeta < 2\gamma $ and $\kappa^2 \zeta>2\gamma$ considered in subsection C, for the matter-dominated universe ($\sigma=0$). There seems to be no simple reason why this should be so.

\section{V. Concluding Remarks}
\label{V}

We have studied one single flat brane situated at $y=0$ in a five-dimensional Randall-Sundrum setting, requiring the spatial curvature parameter $k$ to be zero. The isotropic fluid assumed to be present on the brane was assumed to satisfy the conventional equation of state $p=(\gamma -1)\rho$, with $\gamma$ a constant. The essential new element in our analysis was the allowance of a constant bulk viscosity $\zeta$ in the fluid. The five-dimensional bulk itself, endowed with a five-dimensional cosmological constant $\Lambda$, was assumed not to contain any fluid (or other field).

The governing equations are the Friedmann equation (\ref{22}), in which $\lambda$ can be looked upon as the effective four-dimensional cosmological constant in the five-dimensional theory, together with the energy conservation equation (\ref{26}). 

In Sec.~IV, the main section of our paper, these governing equations were solved under various approximations. The nonviscous theory, briefly considered in subsection A, reproduced essentially results obtained earlier \cite{binetruy00,brevik00}. The results in subsection B in the limit of low energy, are formally the same as those found earlier in a four-dimensional context \cite{brevik94}.  In subsection C, the expressions given analytically for the scalar expansion $\theta$, are to our knowledge new results. In subsection D we introduced the ansatz of Eq.~(\ref{55}) for the bulk kinematic viscosity (being related to the ansatz made earlier by Murphy \cite{murphy72}), and derived on the basis of this the integral expression in Eq.~(\ref{57}) relating cosmological time $t$ to the scale factor $a$. Some simple analytical results were calculated. Moreover, going to the extreme Planck times, our numerical estimates indicated that the viscosity concept may play a physical role in the very early universe.

We emphasize that our theory is built upon the assumption of no flux of energy in the fifth direction, as expressed by Eq.~(\ref{20}).

\subsection{}
\subsubsection{}

\end{document}